\documentclass[10pt,letterpaper,twocolumn,floatfix,ol]{article} 
\usepackage{ol2}
\usepackage[draft,implicit=false]{hyperref}
\usepackage{bm}
\usepackage{color}
\usepackage{graphicx,keyval}
\usepackage{amssymb, amsmath, amsfonts, wasysym}

\begin{document}

\twocolumn[ 

\title{Tunable asymmetric transmission of THz wave through a graphene planar chiral structure}

\author{Junyang Zhao,$^{1}$ Jianfa Zhang,$^{1,*}$ Zhihong Zhu,$^{1,2}$ Xiaodong Yuan,$^{1}$ and Shiqiao Qin$^{1,2}$}

\address{
$^1$College of Optoelectronic Science and Engineering, National University of Defense Technology, Changsha 410073, China\\
$^2$State Key Laboratory of High Performance Computing, National University of Defense Technology, Changsha 410073, China \\

$^*$Corresponding author: jfzhang85@nudt.edu.cn
}

\begin{abstract}
In this letter, we show that asymmetric transmission of circularly polarized waves through a nanostructured planar chiral graphene film can be observed in terahertz range. The asymmetric transmission effect of monatomic layer graphene closely resembles that of metallic planar chiral nanostructures which has previously been demonstrated. And the relative enantiomeric difference in the total transmission varies with the change of graphene's Fermi level. The plasmonic excitation in the graphene nanostructure is the enantiometically sensitive which is asymmetric for opposite propagating directions. This phenomenon will deepen our understanding of light-matter interactions in planar chiral structures and may find applications in polarization-sensitive devices, sensors, detectors and other areas.
\end{abstract}

\ocis{(250.5403) Plasmonics;(160.1190) Anisotropic materials;(310.6628) Subwavelength structures, nanostructures.} 
] 

During the past several years the research on chiral metamaterials have revealed several new electromagnetic phenomena, such as circular dichroism~\cite{decker2007circular,kwon2008optical,valev2009plasmonic} and elliptical dichroism~\cite{zhukovsky2009elliptical} of planar chiral metamaterials, unidirectional or asymmetric transmission of linearly polarized waves in three-dimension chiral metamaterials~\cite{ye2010unidirectional,kang2011asymmetric, menzel2010asymmetric,ye2011unidirectional, mutlu2012diodelike,zhu2012one,shi2013dual,shi2014broadband} or circularly polarized waves in planar chiral structures~\cite{fedotov2006asymmetric, fedotov2007asymmetric,schwanecke2008nanostructured,plum2009planar}. The intriguing optical properties of metallic chiral structures are of great interests both for theoretical research and for practical applications~\cite{yang2009miniature,zhao2012twisted,ye2014circular}. In 2006, it was reported that the first experimental observation of a polarization sensitive transmission effect which is asymmetric with respect to the direction of wave propagation~\cite{fedotov2006asymmetric}. The reported asymmetric phenomenon requires simultaneous presence of planar chirality and anisotropy in the structure which is lossy~\cite{fedotov2007asymmetric}. This new asymmetric transmission effect is different from the symmetric effect analogous to conventional optical activity~\cite{konishi2008observation,kuwata2005giant, plum2009metamaterials,rockstuhl2009optical, singh2010highly} and gyrotropy in 3D-chiral media~\cite{plum2007giant,rogacheva2006giant}. Moreover, its asymmetric transmission behaviour is fundamentally distinct from that with non-reciprocal media~\cite{tymchenko2013faraday} or three-dimensional helical structures~\cite{gansel2009gold,gansel2010gold}. And it has been associated the excitation of enantiomerically sensitive plasmons in the metal nanostructures.

Recently, graphene has raised as a powerful plasmonic material. A monolayer of doped graphene can resemble a thin metal film and support the excitation of surface plasmons in mid-infrared and terahertz (THz) ranges while it is only atomically thick. The exploration of plasmons in graphene nanostructures has leads to the proposition and demonstration of a variety of devices such as tunable polarizers~\cite{yan2012tunable,zhu2014electrically} , metamaterials~\cite{ju2011graphene} and perfect absorbers~\cite{thongrattanasiri2012complete,zhang2014coherent}. Here we show that tunable asymmetric transmission of circularly polarized light can be realized in the THz range with a monolayer graphene patterned with arrays of G-shaped micro-holes which forms a truly planar structure.

\begin{figure}[htbp]
\centerline{\includegraphics[width=85mm]{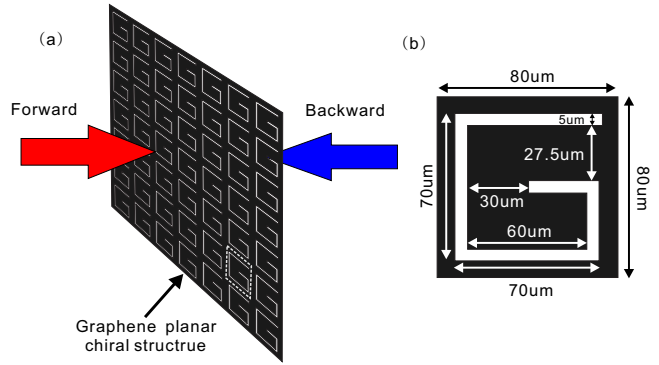}}
\caption{The G-shaped graphene planar chiral structure. (a) Schematics of asymmetric transmission in the arrays of graphene planar chiral structure, one circularly polarized terahertz wave illuminates on the graphene film from opposite sides respectively at normal incidence. (b) A unit cell of the structured graphene film with geometric parameters. The graphene film is patterned with periodical arrays of G-shaped holes. The period is $P = 80 um$ and the width of G model is $5 um$.}
\vspace{-10pt}
\end{figure}

Figure 1 shows the schematic illustration of the graphene planar chiral structure. The graphene is patterned with arrays of G-shape holes with dimensions in the micro-meter scale. One circularly polarized (left or right circular polarization) THz wave normally impinges on the graphene film from forward and backward direction, respectively (see figure 1(a)). Figure 1(b) shows a unit cell of the graphene planar chiral structure with geometric parameters. The black area is graphene and the white area is the removed pattern. The period of the structure is $P = 80~um$, which ensures that the structure does not diffract at the studied spectral range at normal incidence. The side length and width of G-shaped holes are $70~um$ and $5~um$, respectively


\begin{figure}[htbp]
\centerline{\includegraphics[width=84mm]{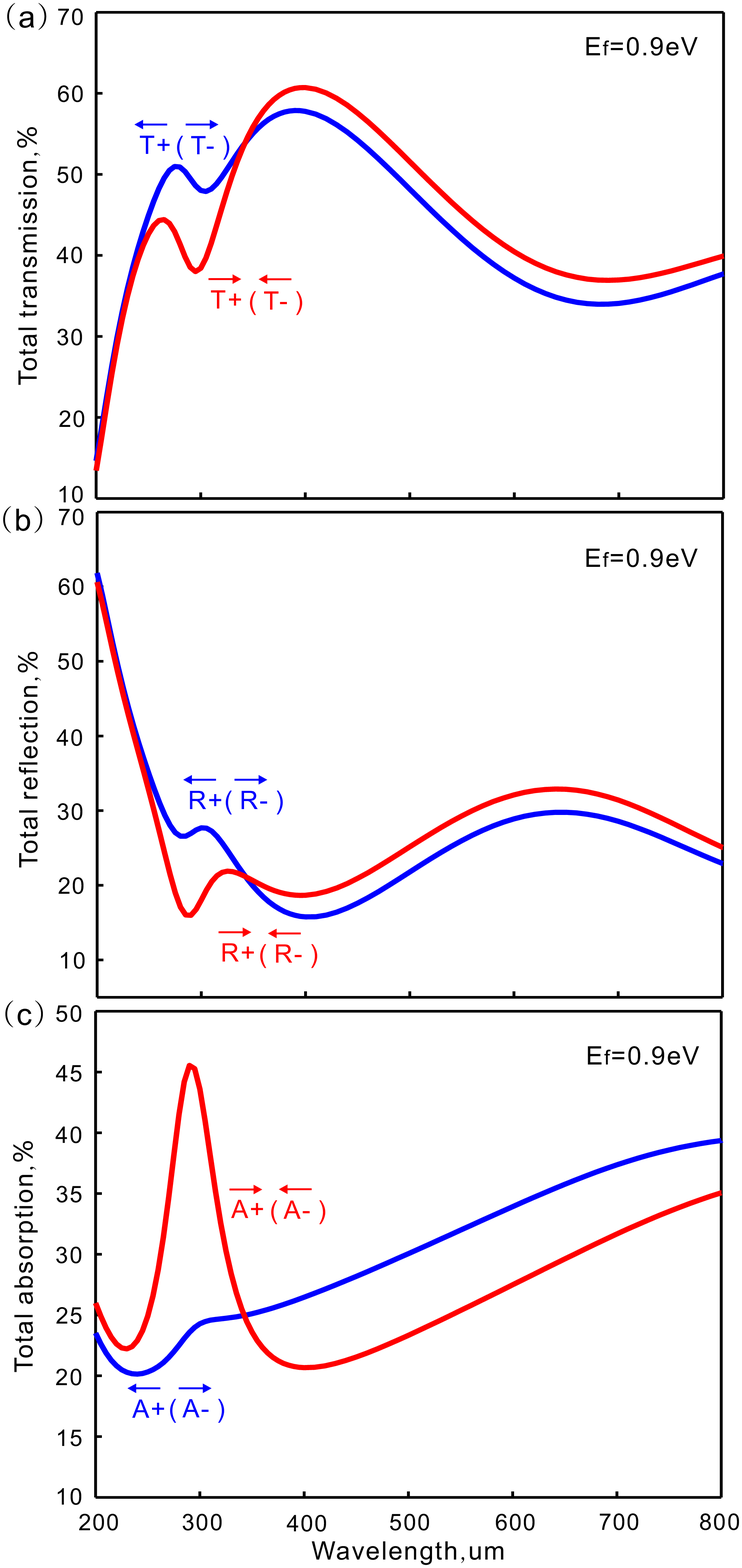}}
\caption{Asymmetric transmission effect in the graphene planar chiral structure. (a) Total transmission, (b) total reflection and (c) total absorption spectra for circularly polarized THz wave are shown. The wave is incident from free space at normal direction. The Fermi level of graphene is to 0.9~eV. Red and blue curves correspond respectively to forward and backward directions of propagation.}
\vspace{-10pt}
\end{figure}

Using a fully three-dimensional finite element numerical simulation method(in Comsol Multi-Physics), with periodic boundary conditions applied to the unit cell, we studied the transmission, reflection and absorption properties of the graphene planar chiral structure in the $200-800 um$ wavelength range for the case of circularly polarized light at normal incidence. The Fermi level of graphene ranges from 0.6~eV to 1.2~eV. The optical conductivity of graphene can be derived within the random-phase approximation (RPA) in the local limit~\cite{falkovsky2007optical,falkovsky2007space}

\begin{equation}
\begin{split}
\sigma_{\omega}=  & \frac{2e^{2} k_{B} T}{\pi \hbar^{2}} \frac{i}{\omega+i\tau^{-1}} ln[2\cosh(\frac{E_{F}}{2k_{B} T})]
\\ & +\frac{e^{2}}{4\hbar}[\frac{1}{2}+ \frac{1}{\pi} arctan(\frac{\hbar \omega-2E_{F}}{2k_{B} T})
\\ & -\frac{i}{2\pi}\ln \frac{(\hbar \omega+2E_{F})^{2}}{(\hbar \omega-2E_{F})^{2}+4(k_{B}T)^{2}}]
\end{split}
\end{equation}
where $k_{B}$ is the Boltzmann constant, $T$ is the temperature, $\omega$ is the frequency of light, $\tau$ is the carrier relaxation lifetime, and $E_{F}$ is the Fermi energy. The first term in Eq.~(1) corresponds to intra-band transitions and the second term is attributed to inter-band transitions. Equation~(1) reduces to the Drude model if we neglect both inter-band transitions and the effect of temperature ($T=0$)
\begin{equation}
\sigma_{\omega}=\frac{e^{2} E_{F}}{\pi \hbar^{2}} \frac{i}{\omega+i\tau^{-1}}
\end{equation}
where $E_{F}$ depends on the concentration of charged doping and $\tau=\mu E_{F}/(ev_{F}^{2})$, where $v_{F}\approx 1\times 10^{6}~m/s$ is the Fermi velocity and $\mu$ is the dc mobility. Here we use a moderate measured mobility $\mu=10000~cm^{2}\cdot V^{-1} \cdot s^{-1}$.

It should be noted that for a certain type of circularly polarized light (i.e., right or left circularly polarized), the transmitted light may change the polarization state and contains both light and left circularly polarized light. The transmission can be presented in terms of a circular transmission $2 \times 2$ matrix $\chi$, whose indexes "+" and "-" denote right (RCP) and left (LCP) circular polarizations, respectively. Matrix $\chi$ will be denoted as $\overrightarrow{\chi}$ for a wave incident on the front side of the structure, while the arrow will be in opposite direction corresponds to the wave incident from the opposite direction. The transmission matrix will be denoted as $\overleftarrow{\chi}$. For instance, the data presented in figure 2a illustrates $\overleftarrow{T_{+}} \neq \overrightarrow{T_{+}}$. For incident RCP wave, the total transmission in the backward direction is given by $\overleftarrow{T_{+}} = \overleftarrow{|\chi_{++}|^{2}} + \overleftarrow{|\chi_{-+}|^{2}}$, while in the opposite direction $\overrightarrow{T_{+}} = \overrightarrow{|\chi_{++}|^{2}} + \overrightarrow{|\chi_{-+}|^{2}}$. Figure 2a shows the total transmission of right circularly polarized wave (defined as the electric vector of the wave coming toward you appears to be rotating counterclockwise) incident from forward and backward directions when the Fermi level of graphene is 0.9~eV. It shows asymmetry depending on the the direction of propagation. According to Ref.~\cite{fedotov2007asymmetric}, $\overleftarrow{|\chi_{++}|^{2}} = \overrightarrow{|\chi_{++}|^{2}}$, while $\overleftarrow{|\chi_{-+}|^{2}} \neq \overrightarrow{|\chi_{-+}|^{2}}$. It means the difference in the total transmission depends on the fact of the asymmetric conversion of transmission for a circular polarized wave into one of opposite handedness. And the total reflection (R) and absorption (A) shows similar asymmetry, as shown in figure 2b and 2c. The results are similar to those of metallic planar chiral metamaterials~\cite{fedotov2007asymmetric}.

\begin{figure}[htbp]
\centerline{\includegraphics[width=84mm]{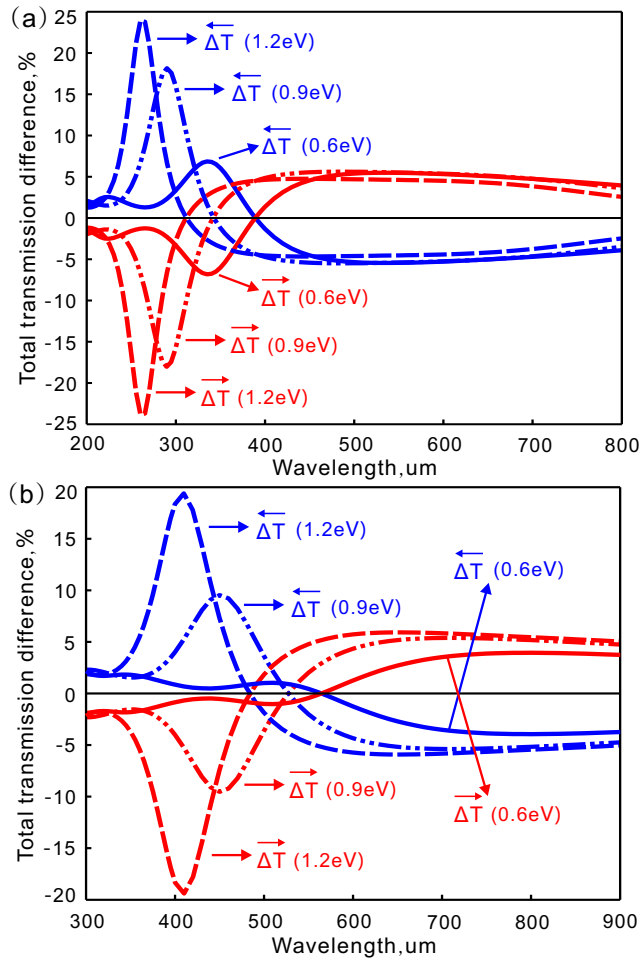}}
\caption{The relative enantiomeric difference in the total transmission. The total transmission difference for patterned graphene planar chiral structure (a) without a substrate and (b) with a substrate of quartz. The Fermi levels of graphene varies from 0.6~eV to 1.2~eV. Red and blue curves correspond respectively to forward and backward directions of propagation.}
\vspace{-10pt}
\end{figure}

Figure 3 shows the relative enantiomeric difference in the total transmission without a substrate and with a substrate of quartz, respectively, for the different Fermi levels. The relative enantiomeric difference in the total transmission is defined as $\overleftarrow{\Delta T} = (\overleftarrow{T_{+}} - \overleftarrow{T_{-}}) / \overleftarrow{T_{+}}$ and $\overrightarrow{\Delta T} = (\overrightarrow{T_{+}} - \overrightarrow{T_{-}}) / \overrightarrow{T_{+}}$. It displays a resonant nature and varies with the Fermi level of graphene. Figure 3a shows the situation for a free-standing graphene as in Figure 2. When Fermi levels are 0.6~eV and 0.9~eV, it reaches its maximum value of about 0.068 and 0.18 at the resonant wavelength of $\lambda=335~um$ and $\lambda=290~um$, respectively. As the Fermi level increases to 1.2~eV, the maximum value increases to 0.24 at the resonant wavelength of about $260 um$.

Figure 3b shows the relative enantiomeric difference in the total transmission obtained in the same graphene planar chiral structure supported on a quartz substrate. The substrate is assumed to be semi-infinite. In our studied Thz range, it can be treated as a transparent material with the refractive index of 1.96 from Ref.~\cite{naftaly2007terahertz}. With the presence of a substrate, the plasmonic resonances of the graphene film redshifts to longer wavelengthes. The results closely resemble those without a substrate presented even thought the relative enantiomeric difference is slightly smaller at the resonances. When Fermi level is 0.6~eV, the difference is unconspicuous. As the Fermi level increases to 1.2~eV, it reaches the maximum value of 0.194 at about $410 um$. Moreover, The sign of $\Delta T$ is changed upon reversal of the propagation direction.

\begin{figure}[htbp]
\centerline{\includegraphics[width=85mm]{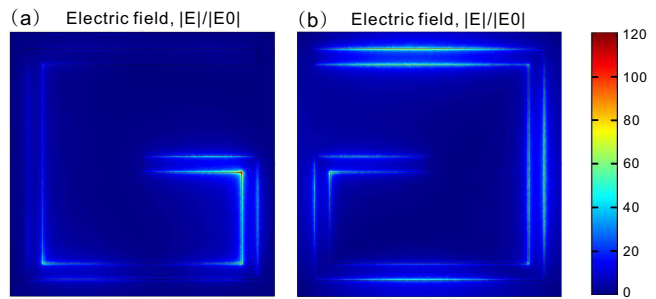}}
\caption{Enantiomerically sensitive plasmonic excitations in the graphene planar chiral structure. The maps show the distribution of normalized electric field in the plane of the structure at the resonance ($\lambda = 290 um$) which the Fermi level of graphene is 0.9~eV. The left circularly polarized light incident on the planar chiral structure from (a) forward and (b) backward sides, respectively.}
\vspace{-10pt}
\end{figure}

Similar to metallic planar chiral structures that have been previously studied, the asymmetric transmission here is attributed to the excitation of enantiomerically sensitive plasmons in the graphene structure. Figure 4 shows normalized electric field intensity maps at the resonance wavelength when the LCP wave propagates in opposite directions. The tFermi level of graphene is 0.9~eV. The figure reveals the strength and location of the plasmonic excitations in the plane of the structure and exhibits striking difference for the THz wave incident from forward and backward directions.

In summary, we have numerically shown the asymmetric transmission effect in graphene planar chiral structures and the results are closely resemble those in metal planar chiral nanostructures. The monatomic layer graphene can be treated as the thinnest material that can be realized and it will deepen our understanding light-matter interactions in planar chiral structures. The relative enantiomeric difference is tunable with the change of Fermi level. The plasmon excitation in the graphene planar chiral structure is enantiometically sensitive and depends on propagating directions of circularly polarized light. The plasmonic dissipation in the graphene structure is also asymmetric for opposite propagating directions. This phenomenon may find applications in polarization-sensitive devices, sensors, detectors and other areas. For example, the enantiometically sensitive excitation of graphene plasmons can be used to detect the circularly polarized light which has significant applications in chemistry and biology.

\emph{Acknowledgment.} This work was supported by National Natural Science Foundation of China [Grant Nos. 11304389 and 61177051] and Ministry of Science and Technology of China [Grant No. 2012CB933501].



\end{document}